\documentclass[]{aastex}
\usepackage{epsf}
\usepackage{graphics}
\usepackage{emulateapj5}

\def\lens{J2004--1349}
\def\ltsima{$\; \buildrel < \over \sim \;$}
\def\lsim{\lower.5ex\hbox{\ltsima}}
\def\gtsima{$\; \buildrel > \over \sim \;$}
\def\gsim{\lower.5ex\hbox{\gtsima}}

\newenvironment{inlinefigure}{
\def\@captype{figure}
\noindent\begin{minipage}{0.999\linewidth}\begin{center}}
{\end{center}\end{minipage}\smallskip}

\newenvironment{inlinetable}{
\def\@captype{table}
\noindent\begin{minipage}{0.999\linewidth}\begin{center}}
{\end{center}\end{minipage}\smallskip}

\medskip

\shorttitle{} 
\shortauthors{Winn, Hall, \& Schechter}
\received{}
\begin{document}

\title{Mass and dust in the disk of a spiral lens galaxy}

\author{Joshua N.\ Winn\altaffilmark{1,2},
Patrick B.\ Hall\altaffilmark{3,4}, and
Paul L.\ Schechter\altaffilmark{5} }


\altaffiltext{1}{Harvard-Smithsonian Center for Astrophysics, 60
Garden st., Cambridge, MA 02138}

\altaffiltext{2}{National Science Foundation Astronomy \& Astrophysics
Postdoctoral Fellow}

\altaffiltext{3}{Princeton University Observatory, Princeton, NJ 08544}

\altaffiltext{4}{Departamento de Astronom\'{\i}a y Astrof\'{\i}sica,
Facultad de F\'{\i}sica, Pontificia Universidad Cat\'{o}lica de Chile,
Casilla 306, Santiago 22, Chile}

\altaffiltext{5}{Department of Physics, Massachusetts Institute of
Technology, 77 Mass.\ ave., Cambridge, MA 02139}

\begin{abstract}
Gravitational lensing is a potentially important probe of spiral
galaxy structure, but only a few cases of lensing by spiral galaxies
are known.  We present Hubble Space Telescope and Magellan
observations of the two-image quasar PMN~J2004--1349, revealing that
the lens galaxy is a spiral galaxy.  One of the quasar images passes
through a spiral arm of the galaxy and suffers 3 magnitudes of
$V$-band extinction.  Using simple lens models, we show that the mass
quadrupole is well-aligned with the observed galaxy disk.  A more
detailed model with components representing the bulge and disk gives a
bulge-to-disk mass ratio of $0.16\pm 0.05$.  The addition of a
spherical dark halo, tailored to produce an overall flat rotation
curve, does not change this conclusion.
\end{abstract}


\section{Introduction}
\label{sec:intro}

A general property of spiral galaxies is that they have three distinct
mass components: a bulge, a disk, and an extended halo of dark matter.
The bulge and disk can be detected and measured in optical images.
The evidence for dark matter comes from observations of luminous
tracers, such as the dynamics of galactic satellites, tidal tails, and
globular clusters; and, most famously, \ion{H}{1} and stellar rotation
curves (as reviewed recently by Sofue \& Rubin 2001 and Combes 2002).
Instead of exhibiting a Keplerian fall-off, the rotation curves are
flat, or nearly so, at a distance of many optical scale lengths from
the galaxy center.

Where the galaxy light is still appreciable, it is not clear what
fraction of the total gravitational force that produces the rotation
curve is provided by the luminous matter, or, equivalently, what is
the shape and density profile of the dark halo.  Even the extreme
hypothesis that the inner rotation curve is produced by the luminous
disk alone (a ``maximum disk''; Van~Albada \& Sancisi 1986) is still
the subject of debate (see, {\it e.g.}, Sackett 1997, Corteau \& Rix
1999, Palunas \& Williams 2000). Dark haloes are traditionally
visualized as spherical, although there is evidence that they are
flattened in the direction perpendicular to the disk, with density
distributions that have a three-dimensional axis ratio $\sim$0.5 on
scales of $\sim$15 kiloparsecs (see Sackett 1999 and references
therein).

Gravitational lensing has long been recognized as a valuable tool for
studying the mass distribution of galaxies, and of spiral galaxies in
particular (see, {\it e.g.}, Maller, Flores, \& Primack 1997, Keeton
\& Kochanek 1998, Koopmans, de~Bruyn, \& Jackson 1998, M\"{o}ller \&
Blain 1998). The image positions and magnifications of a strongly
lensed quasar depend on the projected mass within the radius bounded
by the quasar images, which is typically comparable to the optical
radius of the lens galaxy.  Furthermore, gravitational deflection
depends on the total intervening mass, regardless of its luminosity or
internal dynamics.

Unfortunately, few spiral galaxy lenses are known. Of the present
sample of approximately 80 multiple-image quasars, only 4 are
confidently known to be produced by spiral lenses.  Selection effects
favor the discovery of elliptical lenses over spiral lenses: spirals
have a smaller multiple-image cross section, spirals produce multiple
images with smaller angular separations, and spirals contain large
amounts of dust that can extinguish one or more images (see, {\it
e.g.}, Wang \& Turner 1997; Perna, Loeb, \& Bartelmann 1997; Keeton \&
Kochanek 1998; Bartelmann \& Loeb 1998).

Two of the 4 spiral lenses described previously are poorly suited for
mass modeling. The spiral lens galaxy of B0218+357 (Patnaik {\it et
al.} 1993) is so crowded by the bright quasar images that even its
position has not yet been measured accurately, despite imaging with
the Hubble Space Telescope (Leh\`{a}r {\it et al.} 2000). Likewise,
there appears to be a spiral lens in PKS~1830--211 (Pramesh Rao \&
Subrahmanyan 1988) but the field is so crowded with stars that two
very different conclusions about the lensing scenario have been drawn
from the same data (Courbin {\it et al.} 2002, Winn {\it et al.}
2002).

The ``Einstein cross'' Q2237+0305 is produced by a spiral galaxy
(Huchra {\it et al.} 1985), and is also unusual in another way: the
galaxy has the lowest redshift ($z=0.0394$) of all known lens
galaxies.  This causes the quasar images to appear very close to the
galaxy center ($<$1~kpc), where they are sensitive mainly to the mass
in the bulge rather than the disk or halo.  Nevertheless, some
interesting information about mass on larger scales has been
obtained. For example, Schmidt, Webster, \& Lewis (1998) estimated the
mass of the bar that is seen in optical images, from its shearing
effect on the image configuration.  Trott \& Webster (2002) argued
that the disk is sub-maximum using model constraints from both the
quasar image configuration and \ion{H}{1} rotation measurements at
larger galactocentric distances.

The other well-studied case of spiral lensing is B1600+434 (Jackson
{\it et al.} 1995, Jaunsen \& Hjorth 1997), in which two quasar images
bracket the nearly edge-on disk of the lens galaxy.  Maller, Flores,
\& Primack (1997) presented the first models for this system
consisting of both a halo and a disk, which were extended by Maller
{\it et al.} 2000 to determine the allowed combinations of disk mass
and halo ellipticity.  Koopmans, de~Bruyn, \& Jackson (1998) achieved
a similar goal using models with disk, bulge, and halo
components. However, this system has two undesirable properties for
mass modeling: the lens galaxy has a central dust lane that makes its
position hard to measure accurately, and there is a massive
neighboring galaxy that adds complexity and uncertainty to the models.
In addition, the quasar images are nearly collinear with the galaxy
center, an accidental symmetry that makes it difficult to test whether
the overall mass distribution is aligned with the galaxy disk;
previous studies have assumed this is the case.

In this paper we describe a spiral lens system that offers a new
opportunity for mass modeling. The system, PMN~J2004--1349, is a
two-image quasar originally discovered in a radio lens survey (Winn
{\it et al.} 2001). The lens galaxy was identified in ground-based
optical images but the angular resolution of those images was not good
enough to determine the galaxy's position or morphology. The quasar
and lens galaxy redshifts are unknown. In \S~\ref{sec:hst} we present
Hubble Space Telescope (HST) optical images showing that the lens
galaxy is a spiral galaxy. The quasar images have very different
optical colors. In \S~\ref{sec:dust} we present Magellan data that
extend the color measurements to near-infrared wavelengths, and we
test whether the color differences are consistent with differential
extinction by dust in the lens galaxy.

In \S~\ref{sec:models} we present mass models.  The lens galaxy does
not have obviously massive neighbors, its position is known with
greater accuracy than the problematic cases mentioned above, and it
does not lie along the line between the quasar images.  Yet despite
these advantages, it is still only a two-image system, and two-image
systems do not provide many constraints on lens models.  Our approach
is to determine what can be learned from the simplest plausible mass
models, and then consider more complicated models.  In
\S~\ref{subsec:powerlaw} we use only the quasar image positions and
fluxes to test whether the mass quadrupole is aligned with the
luminous disk, since this is the first system for which such a test
has been possible. Then in \S~\ref{subsec:constml}, we use the HST
surface photometry to constrain a bulge+disk model, in order to
measure the bulge-to-disk mass ratio. In both
\S\S\ref{subsec:powerlaw} and \ref{subsec:constml}, we briefly
consider the implications of an additional assumption: a flat rotation
curve.  Finally, in \S~\ref{sec:summary} we summarize and discuss
future observations that would provide more constraints on spiral
galaxy structure.

\bigskip
\begin{inlinetable}
\centering
\small
\begin{tabular}{lc}
\hline
\hline
Parameter & Value\\
\hline
R.A.$_{\rm NE} -$ R.A.$_{\rm SW}$     & $981.47\pm 1$~mas\\
decl.$_{\rm NE} -$ decl.$_{\rm SW}$   & $552.34\pm 1$~mas\\
R.A.$_{\rm NE} -$ R.A.$_{\rm Gal}$    & $780\pm 5$~mas   \\
decl.$_{\rm NE} -$ decl.$_{\rm Gal}$  & $247\pm 5$~mas   \\
$\mu_{\rm NE} / \mu_{\rm SW}$         & $1.0\pm 0.1$ \\
\hline
\hline
\end{tabular}
\end{inlinetable}
\smallskip
\noindent{\small Table~1 --- Lens model constraints.}
\medskip
\normalsize

\section{HST observations}
\label{sec:hst}

On 2001~May~21, we observed \lens\ with the HST\footnote{Data from the
{\sc nasa/esa} Hubble Space Telescope (HST) were obtained at the Space
Telescope Science Institute, which is operated by the Association of
Universities for Research in Astronomy, Inc., under {\sc nasa}
contract NAS 5-26555.} and WFPC2 camera (program {\sc id}~9133).
Subsequently, on 2001~June~24, the same field was observed by Beckwith
{\it et al.} ({\sc id}~9267), for a high-redshift supernova search.
All together there were 20 dithered exposures using the F814W filter
(hereafter, $I$), totaling 17.2~ksec; and 4 dithered exposures using
the F555W filter ($V$), totaling 5.1~ksec.  The target was centered in
the PC chip, which has a pixel scale of
$0\farcs0456$~pixel$^{-1}$. The exposures were combined and cosmic
rays rejected using the {\sc drizzle} package (Fruchter \& Hook 2001)
and other standard {\sc iraf}\footnote{The Image Reduction and
Analysis Facility ({\sc iraf}) is a software package developed and
distributed by the National Optical Astronomical Observatory, which is
operated by {\sc aura}, under cooperative agreement with the National
Science Foundation.} procedures.

\begin{figure*}[t]
\begin{center}
  \leavevmode
\hbox{%
  \epsfxsize=7in
  \epsffile{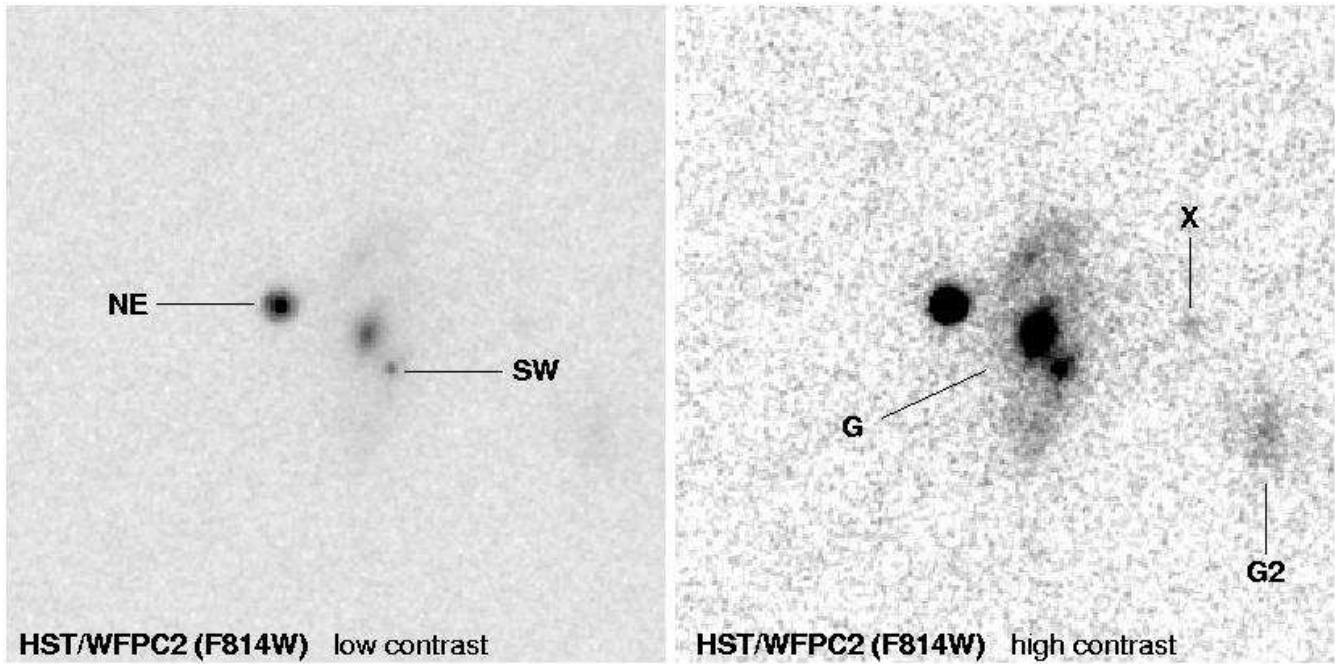}}
\end{center}
\caption{ HST Image of \lens. North is up and east is left. The
quasars are separated by $1\farcs13$.}
\label{fig:hst}
\end{figure*}

Figure 1 shows the final $I$-band image, at low contrast (left panel)
and high contrast (right panel).  The quasars are labeled NE and SW.
The lens galaxy, G, is a spiral galaxy. It has a prominent bulge and a
disk extending nearly north--south. Two loosely wound spiral arms are
visible, spiralling clockwise as they emerge from the eastern and
western sides of the bulge and extend north and south. The spiral arms
are more obvious in the image shown in Figure~3, in which the quasars
and an elliptically symmetric galaxy model have been subtracted using
a procedure described below. The SW quasar passes through the southern
spiral arm, close to the bulge. The $I$-band images also revealed a
faint neighboring galaxy, G2, and another object, X. Object X seems
extended but it is too faint to rule out the possibility that it is a
foreground star.  In $V$-band, the quasars and the lens galaxy bulge
were detected, but with a lower signal-to-noise ratio.  From the HST
images we obtained the following information:

{\it Relative positions.} We measured the relative positions of the
quasars and lens galaxy with point-spread-function (PSF) fitting,
using a nearby star as an empirical PSF (star \#3 from the finding
chart of Winn {\it et al.} 2001).  We modeled the quasars as point
sources, and the galaxy bulge as an elliptical exponential profile
convolved with the PSF. Errors were estimated by the variance in
results of fits to three sub-images, each of which was constructed
from only one-third of the data. The quasar separation agreed with the
more precise value determined by Winn {\it et al.} (2001) using
very-long-baseline interferometry.  The lens galaxy position is given
in Table~1 along with other constraints used in lens models
(\S~\ref{sec:models}). The result did not change significantly when
the lens galaxy was modeled with an elliptical Gaussian profile or a
de~Vaucouleurs profile.

{\it Flux ratio.} The quasar flux ratio (NE/SW) is 8:1 in $I$-band and
20:1 in $V$-band, as compared to 1:1 at radio wavelengths. The SW
quasar image is significantly redder than the NE quasar image. In
\S~\ref{sec:dust} we describe data that extend the measurements to
near-infrared wavelengths and discuss the interpretation.

{\it Position angle and axis ratio of the disk.} By overplotting
ellipses on the $I$-band image, we visually measured the position
angle of the galaxy disk, $\theta_l = -6\fdg5\pm 3\fdg5$, and its
projected axis ratio, $q_l = 0.41\pm 0.05$. We use the subscript $l$
to distinguish properties of the luminosity distribution from
properties of the mass distribution, which we will denote with the
subscript $m$.  Assuming the galaxy is intrinsically a flat circular
disk, the inclination is given by $i=\cos^{-1}q_l = 66\arcdeg\pm
3\arcdeg$.

{\it Standard magnitudes.} To measure the galaxy flux, we used a
$1\farcs5\times 2\farcs5$ rectangular aperture centered on the bulge,
after subtracting the quasars. Table~2 gives the magnitudes, using the
Dolphin (2000) zero points\footnote{As updated at {\tt
http://www.noao.edu/staff/dolphin/wfpc2\_calib/} and corrected to
infinite aperture.} of 21.654 for $I$ and 22.551 for $V$. The total
magnitudes agree with the ground-based photometry of Winn {\it et al.}
(2001), but the magnitudes of the individual components do not
agree. This is because of the much poorer angular resolution of the
ground-based images. In particular, what was previously identified as
the SW quasar in the ground-based images is now known to be mainly
light from the lens galaxy bulge.

\bigskip
\begin{inlinetable}
\centering
\small
\begin{tabular}{lcc}
\hline
\hline
Component & F814W $\approx I$ & F555W $\approx V$ \\
\hline
NE    & $22.06\pm 0.02$ & $24.54\pm 0.04$ \\
SW    & $24.56\pm 0.11$ & $27.79\pm 0.41$ \\
G     & $21.62\pm 0.03$ & $24.30\pm 0.16$ \\
\hline
\hline
\end{tabular}
\end{inlinetable}

\medskip
\noindent{\small Table 2 --- HST photometry.  Error estimates
represent statistical error only, and do not include the zero point
error of $\approx$0.05~mag or CTE effects.}
\medskip
\normalsize

{\it Surface brightness profile.} Figure~2 shows the surface
brightness of the galaxy averaged over elliptical contours.  We
experimented with several parametric fits to the surface brightness
profile, and found the best results for a model consisting of the sum
of two exponential profiles with scale lengths $0\farcs064$ and
$0\farcs65$.\footnote{We note that this decomposition is not unique;
for example, a poorer but still reasonable fit can also be achieved
with a de~Vaucouleurs bulge ($R_{\rm eff}=0\farcs2$) instead of an
exponential bulge.} The solid line shows the profile of the model
after convolving with the PSF. The dashed and dotted lines show the
contributions from each of the two components. The model is a good fit
except for the ``bump'' at a semi-major axis of $0\farcs8$ due to the
spiral arms. The arms are obvious in Figure~3, which shows the
residual image after subtracting the quasars and galaxy model.  In
this model, the bulge-to-disk flux ratio is $0.33\pm 0.07$, where the
disk flux includes the total residual flux within the rectangular
aperture described above.

\begin{figure*}[t]
\begin{center}
  \leavevmode
\hbox{%
  \epsfxsize=3.5in
  \epsffile{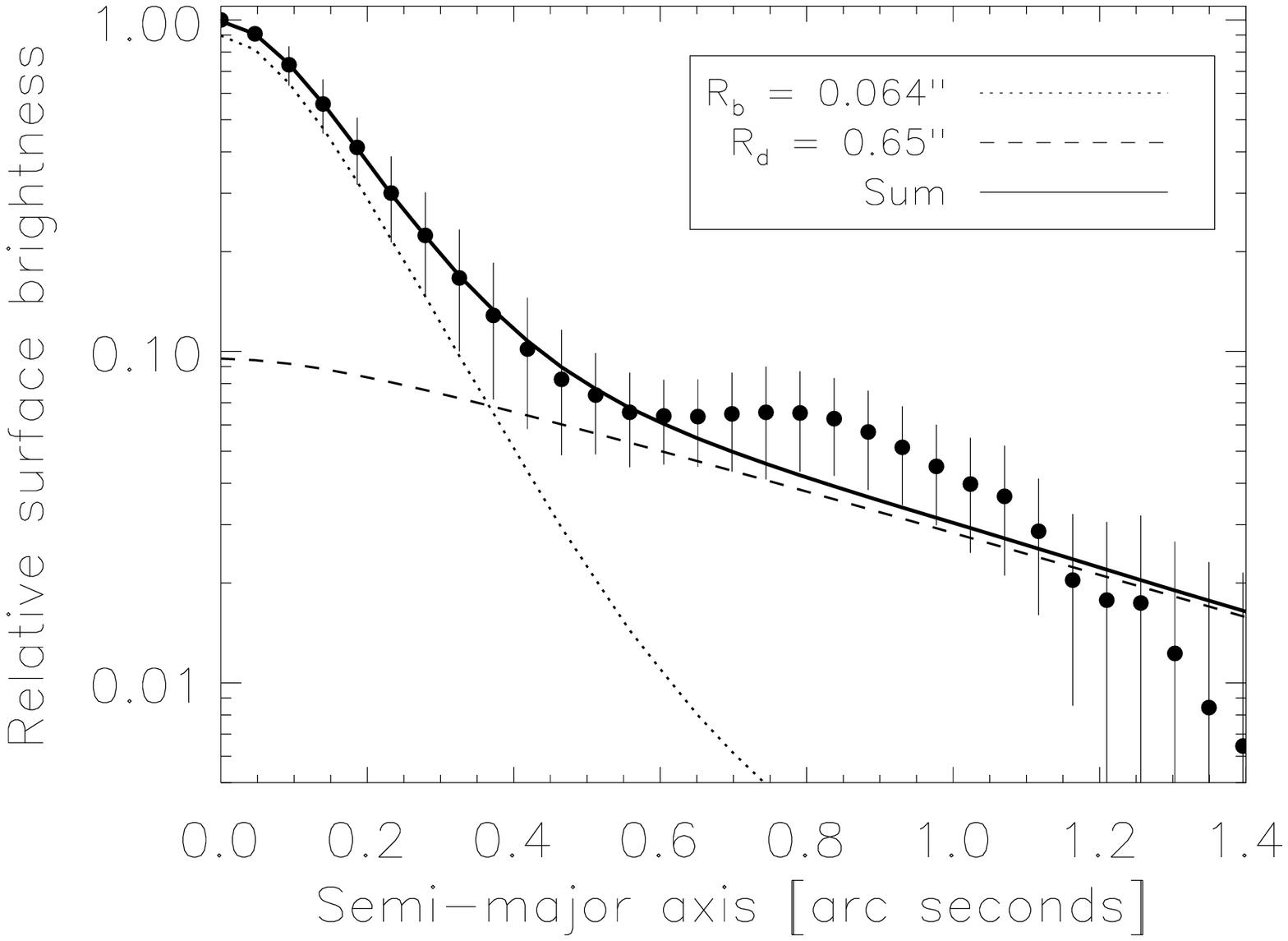}
  \epsfxsize=3.5in
  \epsffile{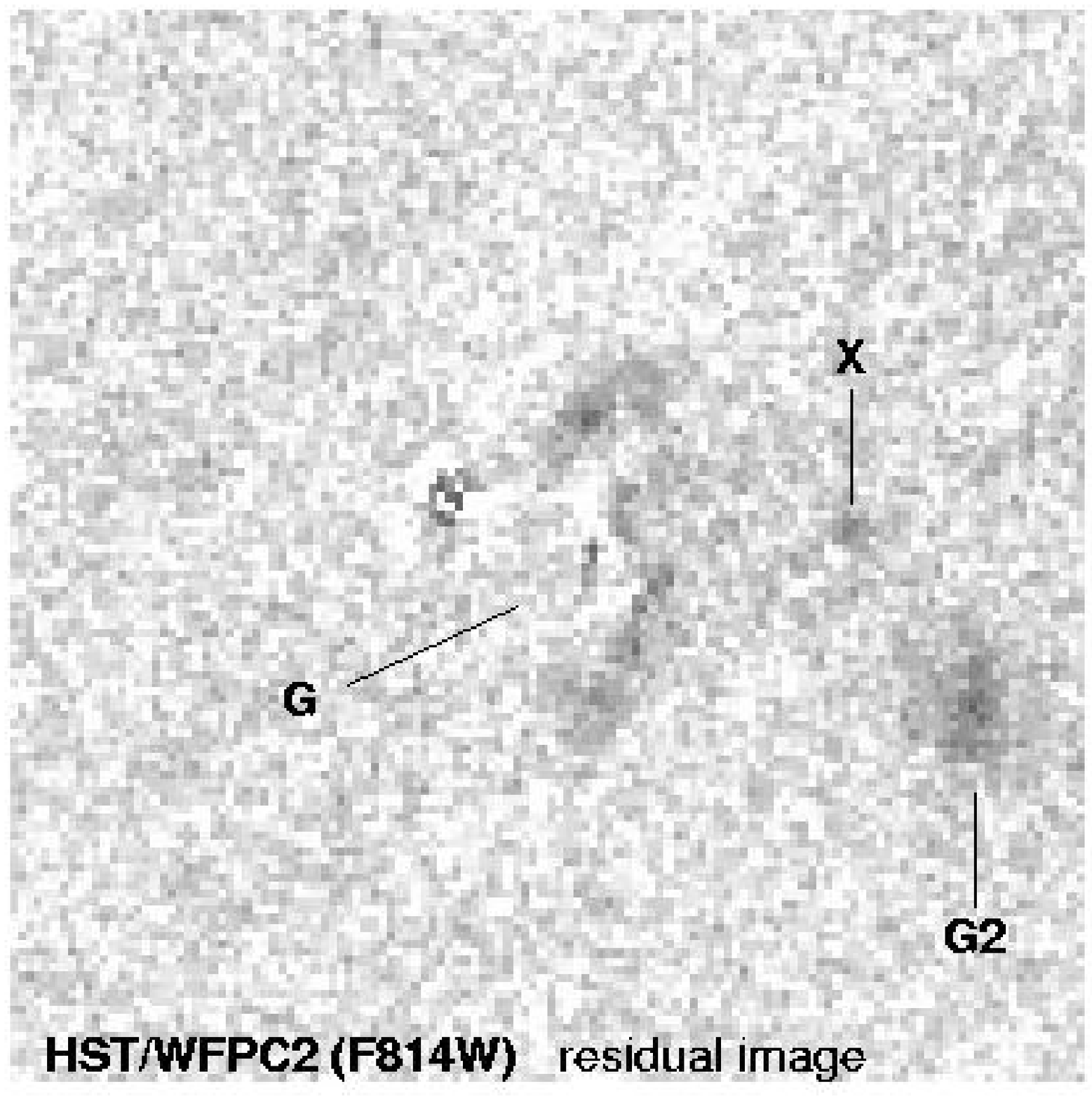}}
\end{center}
\caption{(left) Surface brightness of the lens galaxy, averaged over ellipses
with position angle $\theta_l=-6\fdg5$ and axis ratio
$q_l=0.41$. Error bars show the variance of the points used to compute
the average. The lines show the corresponding surface brightness
profile of the components in the model image.} \label{fig:surfbrt}
\end{figure*}
\begin{figure*}
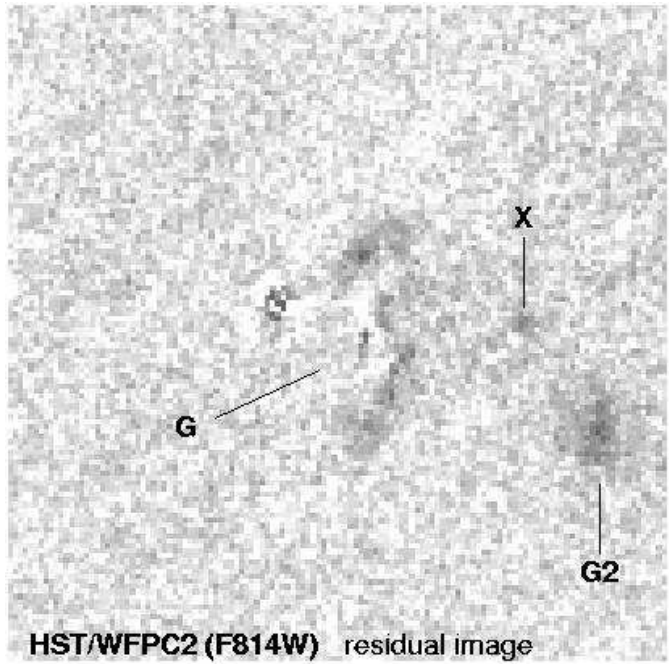

\caption{(right) Residual $I$-band image (=~data~$-$~quasar model~$-$~galaxy
model).  } \label{fig:galaxy}
\end{figure*}

\section{Differential extinction}
\label{sec:dust}

The HST images showed that the SW quasar image is redder than the NE
quasar image.  This trend was confirmed with near-infrared images
obtained on 2002~June~29 with the Baade 6.5m (Magellan I) telescope
and Classic-Cam, a 256$^2$ HgCdTe imager with $0\farcs115$ pixels. We
obtained 52.5 minutes' integration in $J_s$ and 44 in $K_s$, both in
$0\farcs4$ seeing and non-photometric conditions.  The data were
reduced in the standard fashion.  We measured the quasar flux ratio
using the same PSF-fitting procedure that was used for the HST images,
except that in this case we locked the relative positions (and
structural parameters of the galaxy) at the HST-derived values and
allowed only the fluxes to vary.  The uncertainty was estimated by the
spread in results obtained for different choices of the lens galaxy
profile.  Table~3 gives the quasar flux ratio at optical and
near-infrared wavelengths.

\bigskip
\begin{inlinetable}
\centering
\small
\begin{tabular}{lccc}
\hline
\hline
Band & Central wavelength ($\mu$m) & Flux ratio (NE/SW) \\
\hline
$V$     & $0.555$ & $20   \pm 8$ \\
$I$     & $0.814$ & $8.3  \pm 0.7$ \\
$J_s$   & $1.24$  & $2.7  \pm 0.4$ \\
$K_s$   & $2.16$  & $1.44 \pm 0.09$ \\
\hline
\hline
\end{tabular}
\end{inlinetable}
\smallskip
\noindent{\small Table~3 --- Quasar flux ratio.}
\medskip
\normalsize

Because gravitational lensing does not alter the wavelength of
photons, the different colors of NE and SW require explanation. The
most likely explanation is that SW is being reddened by dust as its
light passes through the spiral arm of the lens galaxy. Differential
reddening has been observed in many lens systems (Nadeau {\it et al.}
1991, Falco {\it et al.} 1999), including the spiral lens B1600+434
(Jaunsen \& Hjorth 2002). The main alternative explanations are
microlensing and intrinsic variability.  Microlensing causes color
changes if the angular size of the quasar is unresolved (which is
always the case at optical wavelengths) and varies with wavelength,
because microlensing magnification depends on source size (Wambsganss
\& Paczynski 1991; see Wucknitz {\it et al.} 2003 for a recent
example). Intrinsic variability can also cause color changes when the
time delay is longer than the time scale of variations and the degree
of variability depends on wavelength.  But neither of these
alternative hypotheses is expected to produce color differences as
large as observed here, and neither one would naturally explain why SW
is systematically redder than NE.

To verify that the dust explanation is reasonable, we compare the
wavelength-dependence of the quasar flux ratio, $f_\lambda({\rm
NE})/f_\lambda({\rm SW})$, with what would be expected from dust
reddening. For simplicity we assume that NE is unaffected by dust, and
that intrinsic variability is negligible.  Because the radio flux
density ratio is 1:1, any differences in the observed magnitudes
$m_\lambda$ should be due to extinction $A_\lambda$:
\begin{equation}
m_\lambda({\rm SW}) - m_\lambda({\rm NE}) =
   -2.5\log_{10} \frac{f_\lambda({\rm SW})}{f_\lambda({\rm NE})} =
   A_\lambda({\rm SW}).
\label{eq:extinction}
\end{equation}
We compare the observations with an extinction law $A_\lambda/A_V$
determined by Cardelli, Clayton, \& Mathis (1989), using the standard
Galactic value $R_V=3.1$, and adjusting the wavelength scale as
appropriate for the lens redshift.  There are 2 adjustable parameters,
$E(B-V)$ and $z_{\rm lens}$, and 4 flux ratio measurements. The best
fit ($\chi^2=1.4$) is achieved for $E(B-V)=0.97$ and $z_{\rm
lens}=0.16$, which are both reasonable values.  The measurements and
the fitted extinction curve are shown in Figure~4.  We conclude that
differential extinction is a good explanation for the color
difference, although it is certainly possible that microlensing and
intrinsic variability also contribute to the color difference.

Following Jean \& Surdej (1998), this can be regarded as a measurement
of the lens redshift, but it is a crude one: the range of lens
redshifts giving $\Delta\chi^2<1$ is $0.03 < z_{\rm lens} < 0.36$.  In
addition, the result depends on the choice of $R_V=3.1$.  Larger
values of $R_V$ favor smaller lens redshifts and smaller values of
$R_V$ allow for large lens redshifts. As an example, a solution with
$R_V=2.1$ and $z_{\rm lens}=0.9$ is also plotted in Figure~4.

\bigskip
\begin{inlinefigure}
\begin{center}
\resizebox{\textwidth}{!}{\includegraphics{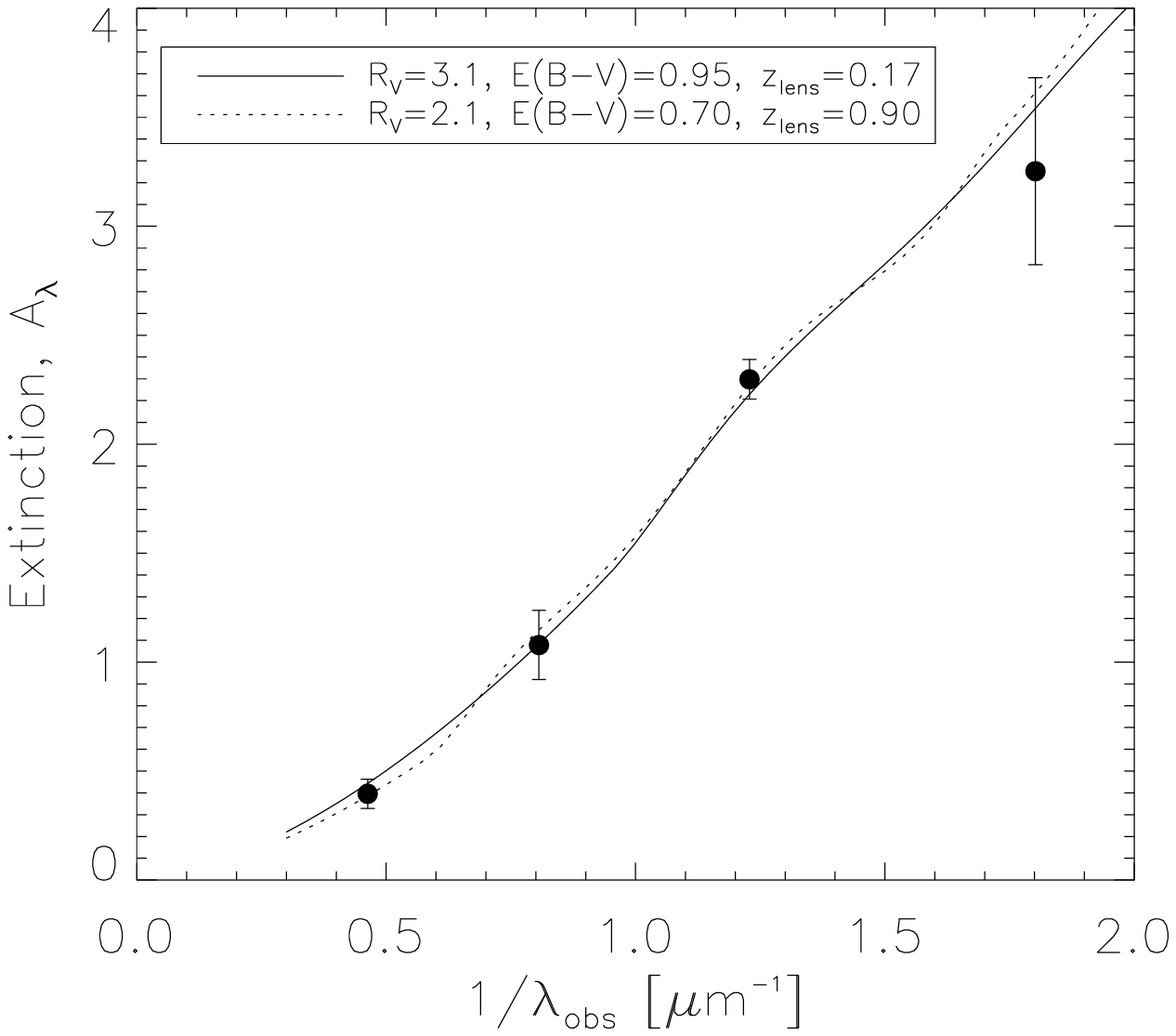}}
\end{center}
\figcaption{ \label{fig:dust} Measured extinction of quasar image SW
(see Eq.~\ref{eq:extinction}) compared to two model extinction laws
from Cardelli, Clayton, \& Mathis (1989). }
\end{inlinefigure}

\section{Lens models}
\label{sec:models}

\subsection{Power-law models}
\label{subsec:powerlaw}

We consider lens models constrained by the image positions and
magnification ratio (as approximated by the radio flux density ratio)
given in Table~1.  The uncertainties in the quasar positions and
magnification ratio were enlarged from the observational uncertainties
to 1~milliarcsecond and 10\%, respectively, to account for systematic
effects due to mass substructure (see, {\it e.g.}, Mao \& Schneider
1998, Metcalf \& Zhao 2002, Dalal \& Kochanek 2002).  With only these
constraints, we are restricted to simple lens models.  Circular models
are too simple because the lens center does not lie along the line
joining the quasars.  We therefore ask: what is the required
ellipticity and position angle of the mass distribution?  From the
qualitative theory presented by Saha \& Williams (2003), we expect the
answer to be fairly model-independent.

We used the ``power-law'' family of mass models,
\begin{equation}
\kappa(\xi) = \frac{1}{2} \left( \frac{b}{\xi} \right) ^{2-\alpha},
\end{equation}
where $\kappa$ is the projected surface density (in units of
$\Sigma_{\rm crit}$, the critical density for strong lensing), $\xi$
is the elliptical coordinate in the image plane, and $b$ is the
Einstein radius.  Implicit in $\xi$ are 4 parameters: the coordinates
of the lens center, the projected axis ratio $q_m$, and the position
angle $\theta_m$. The exponent $\alpha$ determines the rotation curve:
$v(r) \propto r^{\alpha-1}$. The rotation curve is flat for
$\alpha=1$, the isothermal case, and the physically plausible range is
$0 < \alpha < 2$.  With two additional parameters for the source
coordinates, the number of parameters exceeds the number of
constraints by one, leading to a one-dimensional family of models that
fit the data exactly.  We used software written by Keeton (2001) to
optimize the model parameters for a given choice of $q_m$ and
$\theta_m$, by minimizing $\chi^2$ in the source plane (Kayser {\it et
al.} 1989, as modified by Kochanek 1991).

Figure~5 shows the contours of $\chi^2$ in the $(q_m, \theta_m)$
plane.  The line of best-fit solutions is nearly horizontal.  The
direction of the mass quadrupole is therefore well-constrained by the
image configuration even though the radial density profile is not
constrained.  As one moves from left to right along this line, the
mass distribution becomes shallower, with $\alpha$ increasing from
$0\rightarrow 2$. Along the $\chi^2=0$ curve, points are plotted at
intervals of $\Delta\alpha=0.1$ with a symbol size proportional to
$\alpha$. The value of $\alpha$ is indicated explicitly for four
representative models. The caustics and critical curves of those four
models are shown in Figure~6.  Over almost the entire range of
$\alpha$, the position angle of the mass model agrees with the
position angle of the luminous galaxy disk.

\begin{figure*}[t]
\begin{center}
  \leavevmode
\hbox{%
  \epsfxsize=7in
  \epsffile{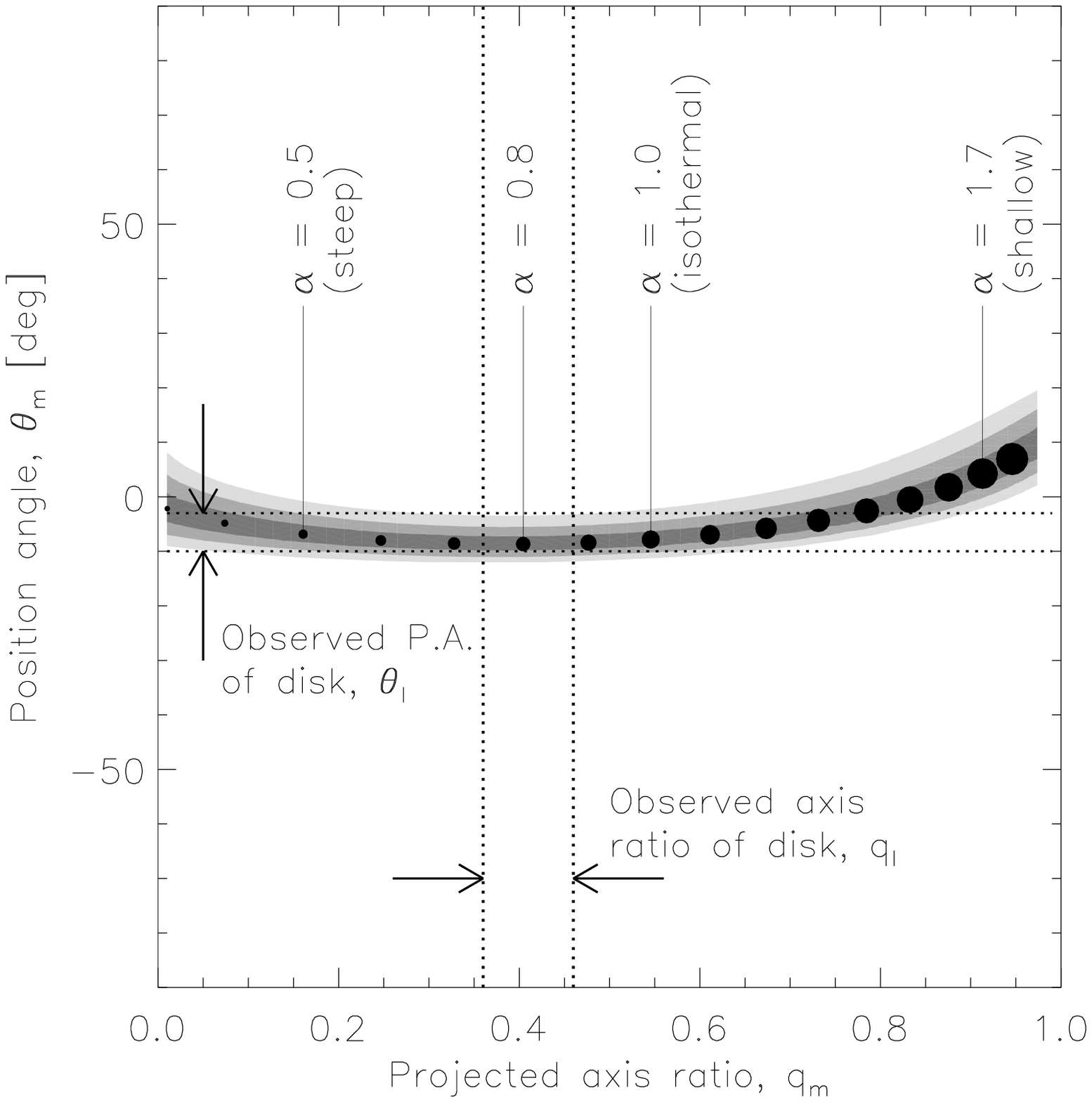}}
\end{center}
\caption{ Allowed power-law models of \lens. Gray levels represent
$\chi^2 < 1$, 4, and 9.  Along the $\chi^2=0$ curve, points are
plotted at intervals of $\Delta\alpha=0.1$ with a symbol size
proportional to $\alpha$. Dotted lines indicate the values determined
from the HST $I$-band image. }
\label{fig:chisq}
\end{figure*}

\begin{figure*}
\begin{center}
  \leavevmode
\hbox{%
  \epsfxsize=7in
  \epsffile{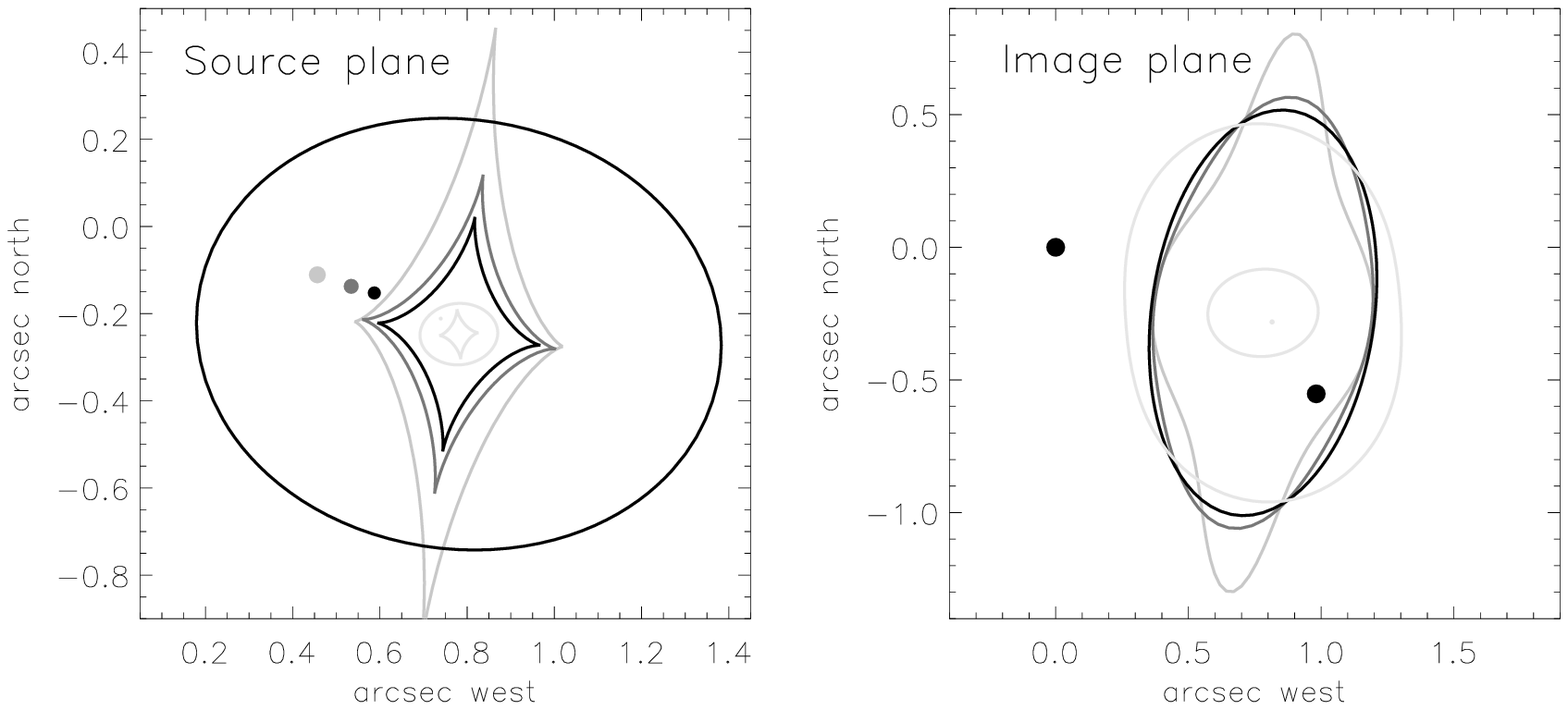}}
\end{center}
\caption{ Lens models of \lens\ for $\alpha=1$ (black), $\alpha=0.8$
(dark gray), $\alpha=0.5$ (light gray), and $\alpha=1.7$ (lightest
gray).  The left panel shows the caustics and source positions. The
right panel shows the critical curves and image positions.  The
innermost ellipse and dot are the radial critical curve and third
image of the $\alpha=1.7$ model.  Source and image fluxes are
proportional to the area of the dots. }
\label{fig:planes}
\end{figure*}

The models for which the mass and light are not quite co-aligned seem
less plausible, on other grounds, than the co-aligned models.  They
are very shallow ($\alpha > 1.49$), corresponding to a rapidly rising
rotation curve at a radius comparable to the disk scale length. They
have a large radial critical curve (which does not exist for
$\alpha\leq 1$) and, within this curve, a third quasar image. These
models are illustrated by the $\alpha=1.7$ case in Figure~6. In that
case, the predicted third image has 7\% of the flux of each bright
image. Current radio maps limit the flux of a third image to $<$2\% of
the flux of each quasar image (Winn {\it et al.} 2001).\footnote{It
should be noted, however, that unmodeled effects---such as a sharp
inner density cusp or central supermassive black hole---might
demagnify the central third image by a large factor without
significantly affecting the other two images (see, {\it e.g.}, Mao,
Witt, \& Koopmans 2001; Keeton 2003).} The shallow models also have
small caustics (and consequently small lensing cross-sections), large
Einstein radii, and large magnifications ($b=3\farcs6$ and $\mu_{\rm
total}=50$, for $\alpha=1.7$).

Therefore, for the most plausible range of mass distributions, the
mass and light are co-aligned.  Another way to state the result is
that we can put only a very weak one-sided limit on the rotation curve
($\alpha<1.49$) by requiring the mass quadrupole to be aligned with
the luminous disk ($\theta_l=\theta_m$).  If we require that the light
and mass agree both in position angle and projected axis ratio
($\theta_l=\theta_m$ and $q_l=q_m$), the result is $\alpha=0.80\pm
0.08$.  However, the latter requirement is too simplistic because it
ignores the contribution of the dark halo, which one might expect to
be rounder than the disk.

Consider, for example, the case of a flat rotation curve: $\alpha=1$.
Figure~6 shows that such a model must be rounder than the observed
disk ($q_m=0.55 > 0.41=q_l$).  As pointed out by Keeton \& Kochanek
(1998), the $\alpha=1$ model can be interpreted as the projection of
either an intrinsically flat Mestel disk, or a singular isothermal
ellipsoid (SIE).  The disagreement of $q_m$ and $q_l$ rules out a pure
Mestel disk, but a SIE halo is allowed.  In terms of the
three-dimensional axis ratio $q_{3m}$, the projected axis ratio $q_m$
of an oblate ellipsoid is
\begin{equation}
q_m^2 = \sqrt{q_{3m}^2\sin^2 i + \cos^2 i},
\end{equation}
giving $q_{3m} = 0.38\pm 0.08$ in this case.  Thus, assuming a flat
rotation curve, a pure-halo model must be very oblate.

\subsection{Models with bulge, disk, and halo}
\label{subsec:constml}

Next, we consider models with separate mass components for the bulge,
disk, and halo.  Based on the HST surface photometry of
\S~\ref{sec:hst}, we described the bulge and disk by exponential
functions,
\begin{equation}
\kappa(\xi) = \kappa_0 \exp(-\xi/R_i),
\end{equation}
fixing the scale lengths of the bulge and disk to be $R_b=0\farcs064$
and $R_d=0\farcs65$.  We required the axis ratio and position angle of
each component to be $0.41\pm 0.05$ and $-6\fdg5\pm 3\fdg5$, to agree
with the HST-measured values, by adding appropriate terms to the
$\chi^2$-function with the software by Keeton (2001).  We allowed the
the central surface density (or, equivalently, the total mass $M$) of
each component to vary.  With only these two components, there is one
degree of freedom.  Even without any dark-matter halo, the model
provides an excellent fit to the data, with $\chi^2_{\rm min} = 0.1$
and $M_b/M_d=0.16\pm 0.03$ for $\Delta\chi^2<1$.

This result and its error bar are internal to our particular
bulge/disk decomposition, but the result is fairly robust, as long as
the bulge is taken to be much more centrally concentrated than the
disk.  For $R_b=0\farcs08$, the maximum value consistent with the HST
image, the result changes only slightly, to $M_b/M_d=0.18$.  The other
extreme, taking the bulge to be a point mass, gives $M_b/M_d=0.14$.
The latter case also demonstrates that the shape of the bulge is not
significant; a perfectly round bulge with $R_b=0\farcs064$ gives
$M_b/M_d=0.15$.  The disk scale length $R_d$ is harder to measure in
the HST image, because of the low surface brightness of the disk and
the patchiness of the arms, but even for $R_d=1\arcsec$ the
bulge-to-disk mass ratio is 0.11.  Hence, insofar as both components
can be approximated by exponential mass distributions, the
bulge-to-disk mass ratio is $0.16\pm 0.05$.

We used the single degree of freedom to investigate how this result
changes with the inclusion of the dark halo.  We described the dark
halo as a spherical isothermal profile with a constant-density core,
\begin{equation}
\kappa(R) = \frac{b}{2} \frac{1}{\sqrt{R^2 + s^2}},
\end{equation}
adding two extra parameters: the Einstein radius $b$ and the core
radius $s$.  We do not consider flattened haloes, because with the
present number of constraints, we would only be able to trace out the
degeneracies between model parameters, which has already been done in
detail by Koopmans, de~Bruyn, \& Jackson (1998) and Maller {\it et
al.} 2000 for the case of B1600+434. Even with a spherical halo, the
number of parameters exceeds the number of constraints by one, and
there is a one-dimensional family of solutions with $\chi^2=0$. We
found all the solutions with $0\arcsec \leq b \leq 2\arcsec$ and, in
each case, determined the net rotation curve and bulge-to-disk mass
ratio.

The spherical halo does not significantly change the derived
bulge-to-disk mass ratio.  The full range of values of $M_b/M_d$ is
0.13--0.17.  The explanation is that when $b$ is large enough to
contribute significantly to the deflection, the optimized core radius
is also large, giving the halo a nearly constant density in the
vicinity of the quasar images. This nearly constant-density sheet
affects only the optimized position and flux of the source. The
relative image positions and fluxes are controlled by the mass
components representing the bulge and disk.

\begin{inlinefigure}
\begin{center}
\resizebox{\textwidth}{!}{\includegraphics{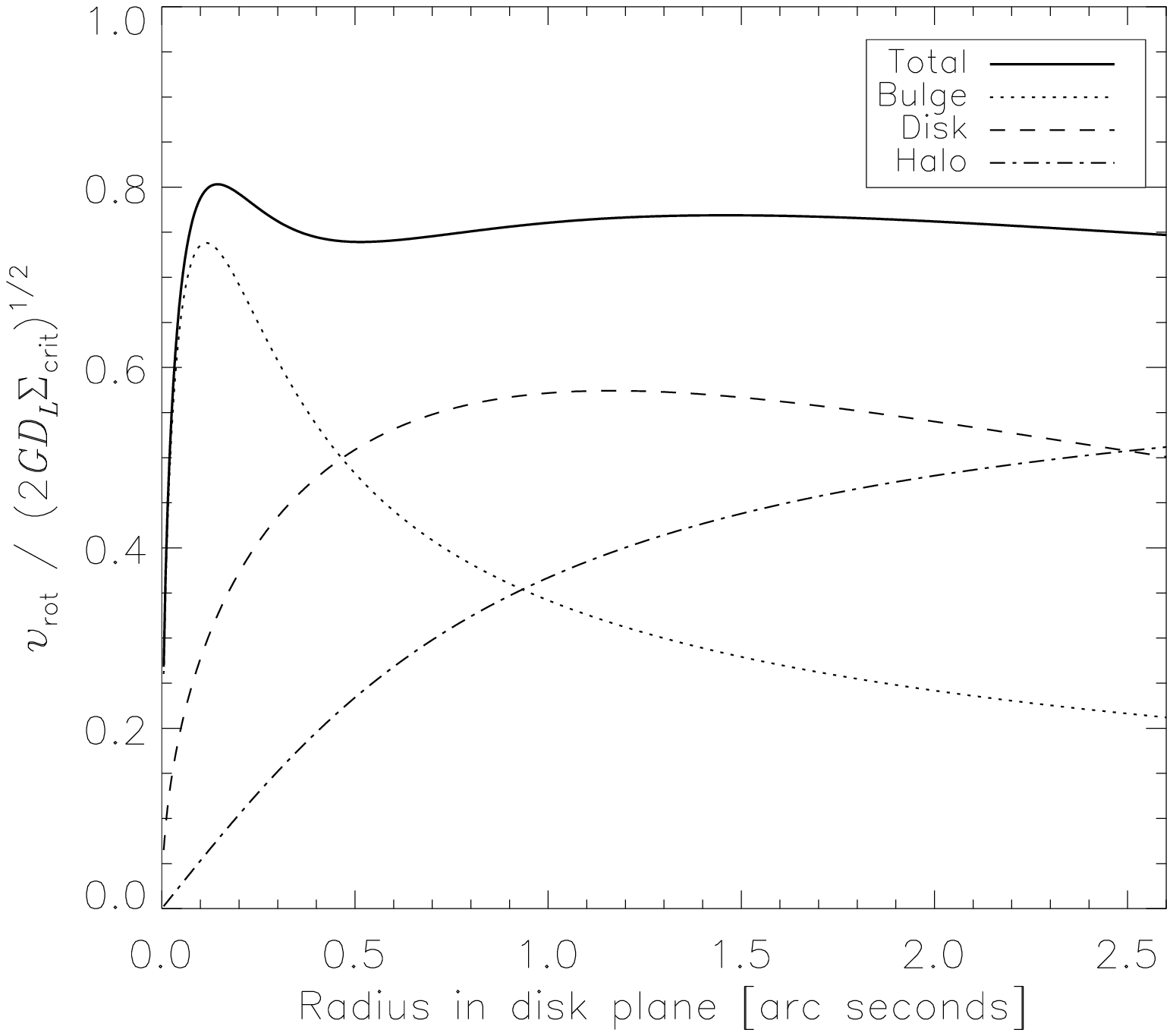}}
\end{center}
\figcaption{ \label{fig:rotcurve} Rotation curve of the
bulge+disk+halo model described in \S~\ref{subsec:constml}. }
\end{inlinefigure}

Although the rotation curve of the lens galaxy has not been measured,
most disk galaxies have fairly flat rotation curves. From all the
disk+bulge+halo models, we determined the one with the flattest
rotation curve (smallest mean-squared deviation from a straight line)
in the range $R_b < r < 4R_d$.  This model has $M_b/M_d=0.16$ and the
rotation curve is plotted in Figure~7.  The individual rotation curves
of the bulge, disk, and halo are also plotted.  Because the redshifts
of the lens and source are unknown, the velocity must be plotted in
dimensionless units $v(R)/\sqrt{2GD_L\Sigma_{\rm crit}}$, where $D_L$
is the angular-diameter distance to the lens, and $\Sigma_{\rm crit}$
is the critical density for strong lensing.

\section{Summary and discussion}
\label{sec:summary}

We have shown that PMN~J2004--1349 has a spiral lens galaxy, making it
one of only a handful of spiral lenses known to date. Uniquely, the
lens has an accurately-measured position, does not have a very massive
neighbor, and is not collinear with the quasar images, which has
allowed us to test whether the mass quadrupole is aligned with the
luminous disk.  The direction of the quadrupole is well constrained
even though the radial density distribution is poorly constrained.  We
found that the mass and light are aligned within a few degrees, except
for very shallow mass distributions that appear physically
implausible.  This had been shown previously for elliptical galaxies,
especially by Keeton, Kochanek, \& Seljak (1997), Keeton, Kochanek, \&
Falco (1998), and Kochanek (2002), with increasingly large samples.
However, this had not been shown before for spiral galaxies, owing to
problems with those few examples of spiral lenses currently known.

This conclusion would be weakened if tidal gravitational forces
(``external shear'') from neighboring masses are producing some of the
observed non-collinearity. One might regard the close alignment of the
light and mass as an argument against a large shear. The only
neighbors visible in the HST images, G2 and X, are faint (with fluxes
$\sim$10\% and $\sim$1\% that of G) and are not positioned along the
disk axis where they would produce the maximum effect.

Using the axis ratio, position angle, and scale lengths measured in
the HST image, we tested a model consisting of a bulge and disk with
constant mass-to-light ratios.  The model successfully reproduces the
image configuration for a bulge-to-disk mass ratio of $0.16\pm 0.05$,
a conclusion that does not change if a spherical dark-matter halo is
added to produce a flat rotation curve.  In $I$-band, the
bulge-to-disk flux ratio was found to be $0.33\pm 0.07$, implying
$(M/L)_b/(M/L)_d = 0.5\pm 0.2$ in $I$-band.  This is a
counter-intuitive result, because one expects disks to contain more
young and massive stars (with smaller mass-to-light ratios) than
bulges.  A flattened halo would reduce the disk mass, but current data
do not provide enough constraints to do more than trace out this
degeneracy.  It would be interesting to assess the stellar populations
of the bulge and disk, and the possible effect of internal extinction
in the disk, using multi-color HST images. The current $V$-band images
are too shallow to provide useful color information for the bulge and
disk separately.

A high priority for future work is spectroscopy of the lens galaxy and
source quasar. Knowledge of the redshifts of lens and source are
essential for computing and interpreting the mass-to-light ratios of
the lens galaxy. Measurement of the circular velocity of the lens
galaxy would break modeling degeneracies between different mass
distributions with the same projected surface density (for analogous
work on elliptical galaxies, see {\it e.g.} Falco {\it et al.} 1997,
Tonry 1998, Romanowsky \& Kochanek 1999, Koopmans \& Treu 2002).

It may also be possible to measure the time delay between flux
variations of the quasar images, which depends sensitively on the
radial mass profile. Or, conversely, it may be possible to measure the
Hubble constant using the Refsdal (1964) method, if the radial density
profile is determined by other means.  Sensitive radio observations
may detect a third quasar image, or lensed radio jets emerging from
the quasar cores, which can discriminate between different radial mass
profiles (Rusin {\it et al.} 2002; Winn, Rusin, \& Kochanek 2003).
Finally, if the sample of spiral lenses can be greatly expanded, it
may be possible to use the ensemble of lens data to place statistical
constraints on spiral galaxy structure, as has been done for
elliptical galaxies (Rusin, Kochanek, \& Keeton 2003).  The enticing
potential of gravitational lensing to probe the mass distribution of
spiral galaxies will depend on the success of at least some of these
proposed observations.

\acknowledgments We are very grateful to Chuck Keeton, for his lens
modeling code; Brian McLeod, for his PSF-fitting code; and Chris
Kochanek and David Rusin, for helpful discussions and comments on the
manuscript. The HST observations were associated with program \#9133,
for which support was provided by {\sc nasa} through a grant from the
Space Telescope Science Institute. We acknowledge the financial
support of NSF grant AST-0104347 (J.N.W.), NSF grant AST-0206010
(P.L.S.) and the Chilean grant FONDECYT/1010981 (P.B.H.).

\end{document}